\documentclass[a4paper,11pt]{paper}
\usepackage[latin1]{inputenc}
\usepackage{graphicx}
\usepackage{multicol}
\usepackage{float}
\usepackage{color}
\usepackage{fancyhdr}
\usepackage{verbatim}
\usepackage{amsmath,amssymb,latexsym}
\usepackage{stmaryrd}
\usepackage{pstricks}
\usepackage{subfigure}
\usepackage{xspace}
\usepackage{natbib}
\bibpunct{(}{)}{,}{a}{,}{,}
\newtheorem{thm}{Theorem}

\newcommand{\bx}{\mathbf{x}}
\newcommand{\bz}{\mathbf{z}}
\newcommand{\bn}{\mathbf{n}}
\newcommand{\ds}{\displaystyle}
\newcommand\dd{\,\text{d}\,}

\newskip\topruleaboveskip%
\newskip\toprulebelowskip%
\newskip\botruleaboveskip%
\newskip\midruleaboveskip%
\newskip\midrulebelowskip%
\def\fullpthrule{\hrule height1.0pt}%
\def\halfpthrule{\hrule height0.5pt}%

\def\toprule{\noalign{\vskip\topruleaboveskip\fullpthrule\vskip\toprulebelowskip}}%
\def\midrule{\noalign{\vskip\midruleaboveskip\halfpthrule\vskip\midrulebelowskip}}%
\def\botrule{\noalign{\vskip\botruleaboveskip\fullpthrule}}%

\title{Exact Bayesian Analysis of Mixtures}
\author{C.P.~Robert$^{1,2}$, K.L.~Mengersen$^3$\\
$^1$Universit\'e Paris-Dauphine, $^2$CREST, Paris, and\\ $^3$Queensland University of Technology, Brisbane}

\begin{document}

\maketitle

\begin{abstract}
In this paper, we show how a complete and exact Bayesian analysis of a
parametric mixture model is possible in some cases when components of the
mixture are taken from exponential families and when conjugate priors are used.
This restricted set-up allows us to show the relevance of the Bayesian approach
as well as to exhibit the limitations of a complete analysis, namely that it is
impossible to conduct this analysis when the sample size is too large, when the
data are not from an exponential family, or when priors that are more complex
than conjugate priors are used. 

\end{abstract}

\section{Introduction}
\label{sec:BAnoMC}

As a warning to the reader, we want to stress from the beginning that this
paper is mostly a formal exercise: to understand how the Bayesian analysis of a
mixture model unravels and automatically exploits the missing data structure of
the model is crucial for grasping the details of simulation methods (not
covered in this paper, see, e.g.,
\citealt{robert:casella:2004,lee:marin:mengersen:robert:2008}) that take full
advantage of the missing structures.  It also allows for a comparison between
exact and approximate techniques when the former are available. While the
relevant references are pointed out in due time, we note here that our paper
builds upon the foundational paper of \citet{fearnhead:2005}.

We thus assume that a sample $\bx=(x_1,\ldots,x_n)$ from the mixture model
\begin{equation}\label{eq:mixexpo}
\sum_{i=1}^k p_i \,h(x)\,\exp\left\{ \theta_i\cdot R(x) - \Psi(\theta_i) \right\}
\end{equation}
is available, where $\theta\cdot R(x)$ denotes the scalar product between the vectors $\theta$
and $R(x)$. We are selecting on purpose the natural representation of an exponential family 
\citep[see, e.g.][Chapter 3]{Robert:2001},
in order to facilitate the subsequent derivation of the posterior distribution.

When the components of the mixture are Poisson $\mathcal{P}(\lambda_i)$ distributions, if
we define $\theta_i=\log\lambda_i$, the Poisson distribution indeed is written as a natural exponential family:
$$
f(x|\theta_i) = (1/x!)\,\exp\left\{ \theta_i\,x-e^{\theta_i} \right\}\,.
$$
For a mixture of multinomial distributions $\mathcal{M}(m;q_{i1},\ldots,q_{iv})$, the natural representation
is given by
$$
f(x|\theta_i) = (m!/x_1! \cdots x_v!)\,\exp\left( x_1 \log q_{i1}+\cdots+x_v \log q_{iv} \right)
$$
and the overall (natural) parameter is thus $\theta_i=(\log q_{i1},\ldots,\log q_{iv})$.

In the normal $\mathcal{N}(\mu_i,\sigma_i^2)$ case, the derivation is more delicate when both
parameters are unknown since
$$
f(x|\theta_i) = \frac{1}{\sqrt{2\pi\sigma_i^2}}\,
	\exp\left( -\frac{\mu_i^2}{2\sigma_i^2} + \frac{\mu_i x}{\sigma_i^2}
	   + \frac{-x^2}{2\sigma_i^2} \right)\,.
$$
In this particular setting, the natural parameterisation is in $\theta_i=(\mu_i/\sigma_i^2,1/\sigma_i^2)$
while the statistic $R(x)=(x,-x^2/2)$ is two-dimensional. The moment cumulant function is then
$\Psi(\theta)=\theta_1^2/\theta_2$.

\section{Formal derivation of the posterior distribution}
\label{sec:ForDerPost}

\subsection{Locally conjugate priors}\label{sub:Locconj}

As described in the standard literature on mixture estimation
\citep{dempster:laird:rubin:1977,maclachlan:peel:2000b,fruhwirth:2006}, the
missing variable decomposition of a mixture likelihood associates each
observation in the sample with one of the $k$ components of the mixture
\eqref{eq:mixexpo}, i.e.  $$ x_i|z_i \sim f(x_i|\theta_{z_i})\,.  $$ Given the
component allocations $\bz$, we end up with a cluster of (sub)samples from
different distributions from the same exponential family. Priors customarily
used for the analysis of these exponential families can therefore be extended
to the mixtures as well.

While conjugate priors do not formally exist for mixtures of exponential
families, we will define {\em locally conjugate priors} as priors that are
conjugate for the completed distribution, that is, for the likelihood
associated with both the observations and the missing data $\bz$. This amounts
to taking regular conjugate priors for the parameters of the different
components and a  conjugate Dirichlet prior on the weights of the mixture,
$$
(p_1,\ldots,p_k) \sim \mathcal{D}(\alpha_1,\ldots,\alpha_k)\,.
$$

When we consider the complete likelihood 
\begin{align*}
L^c(\theta,p|\bx,\bz) &= \prod_{i=1}^n p_{z_i}\,
\exp\left[ \theta_{z_i}\cdot R(x_i) - \Psi(\theta_{z_i}) \right] \\
&= \prod_{j=1}^k p_j^{n_j} \exp\left[ \theta_j\cdot \sum_{z_i=j}
R(x_i) - n_j\Psi(\theta_j) \right] \\
&= \prod_{j=1}^k p_j^{n_j} \exp\left[ \theta_j\cdot S_j - n_j\Psi(\theta_j) \right]\,,
\end{align*}
it is easily seen that we remain within an exponential family since there exists a sufficient
statistic with fixed dimension, $(n_1,S_1,\ldots,n_k,S_k)$. If we use a Dirichlet prior,
$$
\pi(p_1,\ldots,p_k) = \frac{\Gamma(\alpha_1+\ldots+\alpha_k)}
	                   {\Gamma(\alpha_1)\cdots\Gamma(\alpha_k)}
				 p_1^{\alpha_1-1}\cdots p_k^{\alpha_k-1}\,,
$$
on the vector of the weights $(p_1,\ldots,p_k)$ defined on the simplex of $\mathbb{R}^k$,
and (generic) conjugate priors on the $\theta_j$s,
$$
\pi_j(\theta_j] \propto \exp\left[ \theta_j\cdot s_{0j} - \lambda_j\Psi(\theta_j) \right] \,,
$$
the posterior associated with the complete likelihood $L^c(\theta,p|\bx,\bz)$ is then of the same
family as the prior:
\begin{align*}
\pi(\theta,p|\bx,\bz) &\propto \pi(\theta,p)\times L^c(\theta,p|\bx,\bz)\\
&\propto \prod_{j=1}^k p_j^{\alpha_j-1}\,\exp\left[ \theta_j\cdot s_{0j} - \lambda_j\Psi(\theta_j) \right] 
\times p_j^{n_j} \exp\left[ \theta_j\cdot S_j - n_j\Psi(\theta_j) \right] \\
&=\prod_{j=1}^k p_j^{\alpha_j+n_j-1}\,\exp\left[ \theta_j\cdot (s_{0j}+S_j)
- (\lambda_j+n_j)\Psi(\theta_j) \right]\,;
\end{align*}
the parameters of the prior are transformed from $\alpha_j$ to $\alpha_j+n_j$, 
from $s_{0j}$ to $s_{0j}+S_j$ and from $\lambda_j$ into $\lambda_j+n_j$.

For instance, in the case of the Poisson mixture, the conjugate priors are
Gamma $\mathcal{G}(a_j,b_j)$, with corresponding posteriors (for the complete
likelihood), Gamma $\mathcal{G}(a_j+S_j,b_j+n_j)$ distributions, in which $S_j$
denotes the sum of the observations in the $j$th group.

For a mixture of multinomial distributions,
$\mathcal{M}(m;q_{j1},\ldots,q_{jv})$, the conjugate priors are Dirichlet
$\mathcal{D}_v(\beta_{j1},\ldots,\beta_{jv})$ distributions, with corresponding
posteriors $\mathcal{D}_v(\beta_{j1}+s_{j1}+,\ldots,\beta_{jv}+s_{jv})$,
$s_{ju}$ denoting the number of observations from component $j$ in group $u$
$(1\le u\le v)$, with $\sum_u s_{jv}=n_jm$.

In the normal mixture case, the standard conjugate priors are products of normal and inverse gamma
distributions, i.e.
$$
\mu_j|\sigma_j \sim \mathcal{N}(\xi_j,\sigma^2_j/c_j)
\quad\text{and}\quad
\sigma^{-2}_j\sim \mathcal{G}(a_j/2,b_j/2)\,.
$$
Indeed, the corresponding posterior is
$$
\mu_j|\sigma_j \sim \mathcal{N}((c_j\xi_j+n_j\overline{x}_j,\sigma^2_j/(c_j+n_j))
$$
and
$$
\sigma^{-2}_j\sim \mathcal{G}(\{a_j+n_j\}/2,\{b_j+n_j\hat\sigma^2_j	
	+(\overline{x}_j-\xi_j)^2/(c_j^{-1}+n_j^{-1})\})\,,
$$
where $n_j\overline{x}_j$ is the sum of the observations allocated to component $j$ and
$n_j\hat\sigma^2_j$ is the sum of the squares of the differences from $\overline{x}_j$ for
the same group (with the convention that $n_j\hat\sigma^2_j=0$ when $n_j=0$).

\subsection{True posterior distributions}\label{sec:trupod} 
These straightforward derivations do not correspond to the observed likelihood, but
to the completed likelihood. While this may be enough for some simulation
methods like Gibbs sampling \citep[see,
e.g.][]{diebolt:robert:1990a,diebolt:robert:1994}, we need further developments
for obtaining the true posterior distribution.

If we now consider the observed likelihood, it is natural to expand this likelihood as a sum of
completed likelihoods over all possible configurations of the partition space of allocations, that
is, a sum over $k^n$ terms. Except in the very few cases that are processed below, including Poisson
and multinomial mixtures (see Section \ref{ex:fullP}), this sum does not simplify into a smaller
number of terms because there exists no summary statistics. From a Bayesian point of view, the
complexity of the model is therefore truly of magnitude $\text{O}(k^n)$.

The observed likelihood is thus
$$
\sum_{\bz} \prod_{j=1}^k p_j^{n_j} \exp\left\{ \theta_j\cdot S_j - n_j\Psi(\theta_j) \right\}
$$
(with the dependence of $(n_j,S_j)$ upon $\bz$ omitted for notational purposes)
and the associated posterior is, up to a constant,
\begin{align*}
\sum_{\bz} &\prod_{j=1}^k p_j^{n_j+\alpha_j-1} \exp\left\{ \theta_j\cdot (s_{0j}+S_j) 
	- (n_j+\lambda_j)\Psi(\theta_j) \right\} \\
&= \sum_{\bz}\, \omega(\bz)\, \pi(\theta,\mathbf{p}|\bx,\bz)\,,
\end{align*}
where $\omega(\bz)$ is the normalising constant missing in
$$
\prod_{j=1}^k p_j^{n_j+\alpha_j-1} \exp\left\{ \theta_j\cdot (s_{0j}+S_j)
        - (n_j+\lambda_j)\Psi(\theta_j) \right\}
$$
i.e.~
$$
\omega(\bz)\propto \frac{\prod_{j=1}^k \Gamma(n_j+\alpha_j)}{\Gamma(\sum_{j=1}^k \{n_j+\alpha_j\})}\times
\prod_{j=1}^k K(s_{0j}+S_j,n_j+\lambda_j)\,,
$$
if $K(\xi,\delta)$ is the normalising constant of 
$\exp\left\{ \theta_j\cdot \xi - \delta \Psi(\theta_j) \right\}$, i.e.
$$
K(\xi,\delta) = \int \exp\left\{ \theta_j\cdot \xi - \delta \Psi(\theta_j) \right\} \dd\mathbf{\theta}\,.
$$
The posterior $\sum_{\bz}\, \omega(\bz)\, \pi(\theta,\mathbf{p}|\bx,\bz)$ is therefore a mixture of
conjugate posteriors where the parameters of the components as well as the weights can be computed in
closed form! The availability of the posterior does not mean that 
alternative estimates like MAP and MMAP estimates can be computed easily. However,
this is a useful closed form result in the sense that moments can be computed exactly: 
for instance, if there is no label switching problem \citep{stephens:2000b,jasra:holmes:stephens:2005} and,
if the posterior mean is producing meaningful estimates, we have that
$$
\mathbb{E}\left[ \nabla\Psi(\theta_j) | \bx \right] =
\sum_{\bz}\, \omega(\bz)\, \frac{s_{0j}+S_j}{n_j+\lambda_j\,,}
$$
since, for each allocation vector $\bz$, we are in an exponential family set-up where the posterior
mean of the expectation $\Psi(\theta)$ of $R(x)$ is available in closed form. (Obviously, the posterior
mean only makes sense as an estimate for very discriminative priors; see \citealt{jasra:holmes:stephens:2005}.) 
Similarly, estimates of the weights $p_j$ are given by
$$
\mathbb{E}\left[ p_j | \bx \right] =
\sum_{\bz}\, \omega(\bz)\, \frac{n_j+\alpha_j}{n+\alpha_\cdot}\,,
$$
where $\alpha_\cdot = \sum_j \alpha_j$. Therefore, the only computational effort required is the summation over
all partitions. 

This decomposition further allows for a closed form expression of the marginal distributions
of the various parameters of the mixture. For instance, the (marginal) posterior distribution of
$\theta_i$ is given by
$$
\sum_{\bz}\, \omega(\bz)\, \frac{\exp\left[ \theta_j\cdot (s_{0j}+S_j)
        - (n_j+\lambda_j)\Psi(\theta_j) \right]}{K(s_{0j}+S_j,n_j+\lambda_j)}\,.
$$
(Note that, when the hyperparameters $\alpha_j$, $s_{0j}$, and $n_j$ are independent of 
$j$, this posterior distribution is independent of $j$.) Similarly, the posterior 
distribution of the vector $(p_1,\ldots,p_k)$ is equal to
$$
\sum_{\bz}\, \omega(\bz)\, \mathcal{D}(n_1+\alpha_1,\ldots,n_k+\alpha_k)\,.
$$
If $k$ is small and $n$ is large, and when all hyperparameters 
are equal, the posterior should then have $k$ spikes or peaks,
due to the label switching / lack of identifiability phenomenon.

We will now proceed through standard examples.

\subsection{Poisson mixture}\label{ex:fullP}
In the case of a two component Poisson mixture, 
$$
x_1,\ldots,x_n\stackrel{\text{iid}}{\sim}
p\,\mathcal{P}(\lambda_1) + (1-p)\,\mathcal{P}(\lambda_2)\,,
$$
let us assume a uniform prior on $p$ (i.e.~$\alpha_1=\alpha_2=1$) and exponential priors
$\mathcal{E}xp(1)$ and $\mathcal{E}xp(1/10)$ on $\lambda_1$ and $\lambda_2$, respectively.
(The scales are chosen to be fairly different for the purpose of illustration. In a realistic setting,
it would be sensible either to set those scales in terms of the scale of the problem, if known, or
to estimate the global scale following the procedure of \citealt{Mengersen:Robert:1996}.)

The normalising constant is then equal to
\begin{align*}
K(\xi,\delta) &= \int_{-\infty}^\infty \exp\left[ \theta_j \xi 
	- \delta \log(\theta_j) \right] \dd\theta\\
   &= \int_0^\infty \lambda_j^{\xi-1}\,\exp (-\delta\lambda_j)\,\dd\lambda_j \\
   &=  \delta^{-\xi}\,\Gamma(\xi)\,,
\end{align*}
with $s_{01}=1$ and $s_{02}=10$,
and the corresponding posterior is (up to the normalisation of the weights)
\begin{align*}
\sum_{\bz}\, &\frac{\ds \prod_{j=1}^2 \Gamma(n_j+1)\Gamma(1+S_j)\big/(s_{0j}+n_j)^{S_j+1}}{
	\Gamma(2+\sum_{j=1}^2 n_j)}\, \pi(\theta,\mathbf{p}|\bx,\bz) \\
&= \sum_{\bz}\, \frac{\ds \prod_{j=1}^2 n_j!\,S_j! \big/(s_{0j}+n_j)^{S_j+1}}{
	(N+1)!}\, \pi(\theta,\mathbf{p}|\bx,\bz) \\
&\propto \sum_{\bz}\, \prod_{j=1}^2 n_j!\,S_j! \big/(s_{0j}+n_j)^{S_j+1} \, 
	\pi(\theta,\mathbf{p}|\bx,\bz)\,,
\end{align*}
with $\pi(\theta,\mathbf{p}|\bx,\bz))$ corresponding to a Beta $\mathcal{B}e(1+n_j,1+N-n_j)$ 
distribution on $p$ and to a Gamma $\mathcal{G}a(S_j+1,s_{0j}+n_j)$ distribution 
on $\lambda_j$ $(j=1,2)$.

An important feature of this example is that the sum does not need to involve all of the $2^n$
terms, simply because the individual terms in the previous sum factorise in $(n_1,n_2,S_1,S_2)$,
which then acts like a local sufficient statistic. Since $n_2=n-n_1$ and $S_2=\sum x_i - S_1$, the
posterior only requires as many distinct terms as there are distinct values of the pair $(n_1,S_1)$
in the completed sample. For instance, if the sample is $(0,0,0,1,2,2,4)$, the distinct values of
the pair $(n_1,S_1)$ are $(0,0),(1,0),(1,1),(1,2),(1,4),
(2,0),(2,1),(2,2),(2,3),\allowbreak(2,4),(2,5),(2,6),\allowbreak\ldots,(6,5),\allowbreak(6,7),
\allowbreak(6,8),\allowbreak(7,9)$.  There are therefore $41$ distinct terms in the posterior, rather than
$2^8=256$.

The problem of computing the number (or cardinality) $\mu_n(n_1,S_1)$ of terms in the $k^n$ sum with
the same statistic $(n_1,S_1)$ has been tackled by \citet{fearnhead:2005} in that he proposes a
recursive formula for computing $\mu(n_1,S_1)$ in an efficient way, as expressed below for
a $k$ component mixture:

\begin{thm}{{\bf \citep{fearnhead:2005}}}
If $\mathbf{e}_j$ denotes the vector of length $k$ made up of zeros everywhere
except at component $j$ where it is equal to one, if 
$$
\mathbf{n}=(n_1,\ldots,n_k)\,,\quad
\mathbf{n}-\mathbf{e}_j=(n_1,\ldots,n_j-1,\ldots,n_k)\,, \quad\text{and}\quad 
y\mathbf{e}_j=(0,\ldots, y,\ldots,0)\,,
$$
then
$$
\mu_1(\mathbf{e}_j,y\mathbf{e}_j) = 1
\quad\text{and}\quad
\mu_n(\mathbf{n},\mathbf{s})=\sum_{j=1}^k \mu_{n -1}(\mathbf{n}-\mathbf{e}_j,\mathbf{s}-y_n\mathbf{e}_j).
$$
\end{thm}

Therefore, once the $\mu_n(\mathbf{n},\mathbf{s})$ are all computed, 
the posterior can be written
as
$$
\sum_{(n_1,S_1)}\, \mu_n(n_1,S_1) \prod_{j=1}^2 \left[ n_j!\,S_j! \big/(s_{0j}+n_j)^{S_j+1}\right]
\, \pi(\theta,\mathbf{p}|\bx,n_1,S_1)\,,
$$
up to a constant,
since the complete likelihood posterior only depends on the sufficient statistic $(n_1,S_1)$.

Now, the closed-form expression allows for a straightforward representation of the marginals. 
For instance, the marginal in $\lambda_1$ is given by
\begin{align*}
\sum_{\bz}\, &\left( \prod_{j=1}^2 n_j!\,S_j! \big/(s_{0j}+n_j)^{S_j+1}\right)
\,(n_1+1)^{S_1+1}\lambda^{S_1}\,\exp\{-(n_1+1)\lambda_1\} / n_1!\\
&=\sum_{(n_1,S_1)}\mu_n(n_1,S_1) \,\prod_{j=1}^2 n_j!\,S_j! \big/(s_{0j}+n_j)^{S_j+1}
\\
&\qquad\times (n_1+1)^{S_1+1}\lambda_1^{S_1}\,\exp\{-(n_1+1)\lambda_1\} / n_1! \,
\end{align*}
up to a constant, while the marginal in $\lambda_2$ is
$$
\sum_{(n_1,S_1)}\,\mu_n(n_1,S_1) \,\prod_{j=1}^2 \left( n_j!\,S_j! \big/(s_{0j}+n_j)^{S_j+1} \right)
\,(n_2+10)^{S_2+1}\lambda_2^{S_2}\,\exp\{-(n_2+10)\lambda_2\} / n_2!\,
$$
again up to a constant, and the marginal in $p$ is
\begin{align*}
\sum_{(n_1,S_1)}\,\mu_n(n_1,S_1) &\frac{\prod_{j=1}^2 n_j!\,S_j! \big/(s_{0j}+n_j)^{S_j+1}}{
(N+1)!}\,\frac{(N+1)!}{n_1!(N-n_1)!}\,p^{n_1}(1-p)^{N-n_1}\\
&=\sum_{u=0}^N \sum_{S_1;n_1=u} \mu_n(u,S_1)\,
\frac{S_1!(S-S_1)!\,p^u(1-p)^{N-u}}{(u+1)^{S_1+1}(n-u+10)^{S-S_1+1}}\,,
\end{align*}
still up to a constant, if $S$ denotes the sum of all observations.

As pointed out above, another interesting outcome of this closed-form representation is that marginal
likelihoods (or evidences) can also be computed in closed form. The marginal distribution
of $\bx$ is directly related to the unormalised weights $\omega(\bz)=\omega(n_1,S_1)$ in
that
\begin{align*}
m(\bx) &= \sum_\bz \omega(\bz) = \sum_{(n_1,S_1)} \mu_n(n_1,S_1) \omega(n_1,S_1)
       \\&= \sum_{(n_1,S_1)} \mu_n(n_1,S_1)\, \frac{\prod_{j=1}^2 n_j!\,S_j!
	\big/ (s_{0j}+n_j)^{S_j+1}}{(N+1)!}\,,
\end{align*}
up to the product of factorials $1/y_1!\cdots y_n!$ (but this is irrelevant in the
computation of the Bayes factor).

In practice, the derivation of the cardinalities $\mu_n(n_1,S_1)$ can be done 
recursively as in \citet{fearnhead:2005}:
include each observation $y_k$ by updating all the $\mu_{k-1}(n_1,S_1,k-1-n_1,S_2)$s in both
$\mu_k(n_1+1,S_1+y_k,n_2,S_2)$ and $\mu_k(n_1,S_1,n_2+1,S_2+y_k)$,
and then check for duplicates. 
Below is a na\"ive {\sf R} implementation (for
reasonable efficiency, the algorithm should be programmed in a faster language like
{\sf C}.), where \verb+ncomp+ denotes the number of components:
\begin{verbatim}
#Matrix of sufficient statistics, last column is number of occurrences
cardin=matrix(0,ncol=2*ncomp+1,nrow=ncomp)

#Initialisation
for (i in 1:ncomp) cardin[i,((2*i)-1):(2*i)]=c(1,dat[1])
cardin[,2*ncomp+1]=1

#Update
for (i in 2:length(dat)){

    ncard=dim(cardin)[1]
    update=matrix(t(cardin),ncol=2*ncomp+1,nrow=ncomp*ncard,byrow=T)

    for (j in 0:(ncomp-1)){

      update[j*ncard+(1:ncard),(2*j)+1]=
            update[j*ncard+(1:ncard),(2*j)+1]+1
      update[j*ncard+(1:ncard),(2*j)+2]=
            update[j*ncard+(1:ncard),(2*j)+2]+dat[i]
      }

    update=update[do.call(order,data.frame(update)),]
    nu=dim(update)[1]
    #changepoints
    jj=c(1,(2:nu)[apply(abs(update[2:nu,1:(2*ncomp)]-
              update[1:(nu-1),1:(2*ncomp)]),1,sum)>0])
    # duplicates or rather ncomplicates!
    duplicates=(1:nu)[-jj]
    if (length(duplicates)>0){

      for (dife in 1:(ncomp-1)){

        ji=jj[jj+dife<=nu]
        ii=ji[apply(abs(update[ji+dife,1:(2*ncomp)]-
                   update[ji,1:(2*ncomp)]),1,sum)==0]
        if (length(ii)>0)
         update[ii,(2*ncomp)+1]=update[ii,(2*ncomp)+1]+
                            update[ii+dife,(2*ncomp)+1]
      }
      update=update[-duplicates,]
      }

    cardin=update
    }
\end{verbatim}
At the end of this program, all non-empty realisations of the sufficient $(n_1,S_1)$
are available in the two first columns of \verb+cardin+, while the corresponding
$\mu_n(n_1,S_1)$ is provided by the last column.

Once the $\mu_n(n_1,S_1)$'s are available, the corresponding weights can be added as the last
column of \verb+cardin+, i.e.~
\begin{verbatim}
w=log(cardin[,2*ncomp+1])+apply(lfactorial(cardin[,2*(1:ncomp)-1]),1,sum)+
       apply(lfactorial(cardin[,2*(1:ncomp)]),1,sum)-
       apply(log(xi[1:ncomp]+cardin[,2*(1:ncomp)-1])*
        (cardin[,2*(1:ncomp)]+1),1,sum)- sum(lfactorial(dat))
w=exp(w-max(w))
cardin=cbind(cardin,w)
\end{verbatim}
where \verb+xi[j]+ denotes $s_{0j}$.
The marginal posterior on $\lambda_1$ can then be plotted via
\begin{verbatim}
marlam=function(lam,comp=1){
   sum(cardin[,2*(ncomp+1)]*dgamma(lam,shape=cardin[,2*comp]+1,
             rate=cardin[,2*comp-1]+xi[comp]))/sum(cardin[,2*(ncomp+1)])
 }
lalam=seq(.01,1.2*max(dat),le=100)
mamar=apply(as.matrix(lalam),1,marlam,comp=1)
plot(lalam,mamar,type="l",xlab=expression(mu[1]),ylab="",lwd=2)
\end{verbatim}
while the  marginal posterior on $p$ is given through
\begin{verbatim}
marp=function(p,comp=1){
sum(cardin[,2*(ncomp+1)]*dbeta(p,shape1=cardin[,2*comp-1]+1,
      shape2=length(dat)-cardin[,2*comp-1]+1))/sum(cardin[,2*(ncomp+1)])
}
pepe=seq(.01,.99,le=99)
papar=apply(as.matrix(pepe),1,marp)
plot(pepe,papar,type="l",xlab="p",ylab="",lwd=2)
\end{verbatim}

Now, even with this considerable reduction in the complexity of the posterior distribution (to be
compared with $k^n$), the number of terms in the posterior still grows very fast both with $n$ and
with the number of components $k$, as shown through a few simulated examples in Table
\ref{tab:eXplose}.  (The missing items in the table simply took too much time or too much memory on
the local mainframe when using our \verb+R+ program. \citealt{fearnhead:2005} used a specific
\verb+C+ program to overcome this difficulty with larger sample sizes.) The computational
pressure also increases with the range of the data; that is, for a given value of $(k,n)$, the
number of rows in \verb+cardin+ is much larger when the observations are larger, as shown for
instance in the first three rows of Table \ref{tab:eXplose}: a simulated Poisson
$\mathcal{P}(\lambda)$ sample of size $10$ is primarily made up of zeros when $\lambda=.1$ but mostly
takes different values when $\lambda=10$.  The impact on the number of sufficient statistics can be
easily assessed when $k=4$. (Note that the simulated dataset corresponding to $(n,\lambda)=(10,0.1)$
in Table \ref{tab:eXplose} corresponds to a sample only made up of zeros, which explains the
$n+1=11$ values of the sufficient statistic $(n_1,S_1)=(n_1,0)$ when $k=2$.)

\begin{table}\begin{small}
\begin{center}
\begin{tabular}{l  ccc}
\toprule
$(n,\lambda)$ 	& $k=2$ & $k=3$ & $k=4$\\
\midrule
$(10,0.1)$	&  11   &   66    &  286 \\
$(10,1)$	&  52   &  885    & 8160 \\
$(10,10)$	& 166	& 7077    & 120,908\\	         
$(20,0.1)$	&  57   &  231    & 1771 \\
$(20,1)$	& 260   & 20,607  & 566,512\\
$(20,10)$	& 565   & 100,713 &  --- \\
$(30,0.1)$	&  87	&  4060   & 81,000\\
$(30,1)$ 	& 520   & 82,758  &  --- \\
$(30,10)$ 	& 1413  & 637,020 &  --- \\
$(40,0.1)$	& 216   & 13,986  &  --- \\
$(40,1)$        & 789   & 271,296 &  --- \\
$(40,10)$       & 2627  &  ---    &  --- \\
\botrule
\end{tabular}
\caption{\label{tab:eXplose}
Number of pairs $(n_1,S_1)$ for simulated datasets
from a Poisson $\mathcal{P}(\lambda)$ and different
numbers of components. {\em (Missing terms are due 
to excessive computational or storage requirements.)}}
\end{center}
\end{small}\end{table}

An interesting comment one can make about this decomposition of the posterior distribution is that
it may happen that, as already noted in \citet{casella:robert:wells:1999}, a small number of values of the
local sufficient statistic $(n_1,S_1)$ carry most of the posterior weight. Table \ref{tab:cumuweit}
provides some occurrences of this feature, as for instance in the case $(n,\lambda)=(20,10)$.

\begin{table}\begin{center}
\begin{tabular}{l ccc}
\toprule
$(n,\lambda)$   & $k=2$ & $k=3$ & $k=4$\\
\midrule
$(10,1)$ 	& 20/44 & 209/675 & 1219/5760\\
$(10,10)$	& 58/126 & 1292/4641  & 13,247/78,060   \\
$(20,0.1)$	& 38/40 & 346/630 & 1766/6160 \\
$(20,1)$	& 160/196 & 4533/12,819 & 80,925/419,824 \\
$(10,0.1,10,2)$   & 99/314 & 5597/28,206 & --- \\
$(10,1,10,5)$   & 21/625 & 13,981/117,579 & --- \\
$(15,1,15,3)$   & 50/829 & 62,144/211,197 & --- \\ 
$(20,10)$	& 1/580 & 259/103,998 & --- \\
$(30,0.1)$	& 198/466 &  20,854/70,194 & 30,052/44,950\\
$(30,1)$	& 202/512 & 18,048/80,470 & --- \\
$(30,5)$	& 1/1079 & 58,820/366,684 & --- \\
\botrule
\end{tabular}
\caption{\label{tab:cumuweit}
Number of sufficient statistics $(n_i,S_i)$ corresponding to the $99$\%~largest posterior weights/total number of pairs
for datasets simulated either from a Poisson $\mathcal{P}(\lambda)$ or from
a mixture of two Poisson $\mathcal{P}(\lambda_i)$, and different
numbers of components.{\em (Missing terms are due to excessive computational or storage requirements.}}
\end{center}\end{table}

\begin{figure}
\includegraphics[width=\textwidth]{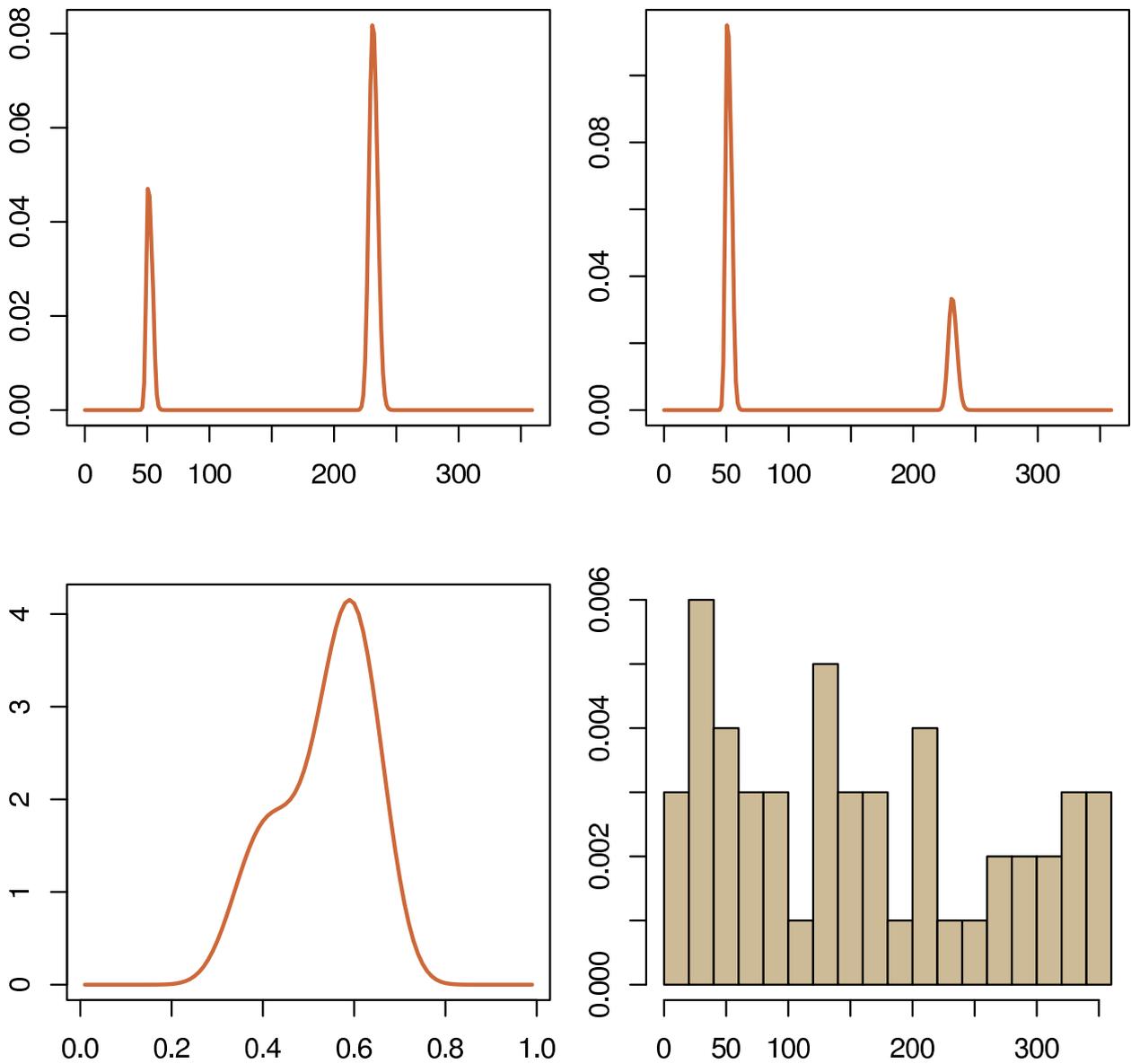}
\caption{
{\em (top left)} Marginal posterior distribution of $\lambda_1$
{\em (top right)} marginal posterior distribution of $\lambda_2$
{\em (bottom left)} marginal posterior distribution of $p$
{\em (bottom right)} histogram of the minnow dataset.
(The prior parameters are $1/100$ and $1/200$ to remain compatible
with the data range.)}\label{fig:topminnow1}
\end{figure}

We now turn to a minnow dataset made of $50$ observations, for which we need a minimal description.
As seen in Figure \ref{fig:topminnow1}, the datapoints take large values, which is a drawback
from a computational point of view since the number of statistics to be registered is much larger than
when all datapoints are small. For this reason, we can only process the mixture model with $k=2$
components.

\begin{figure}
\includegraphics[width=\textwidth]{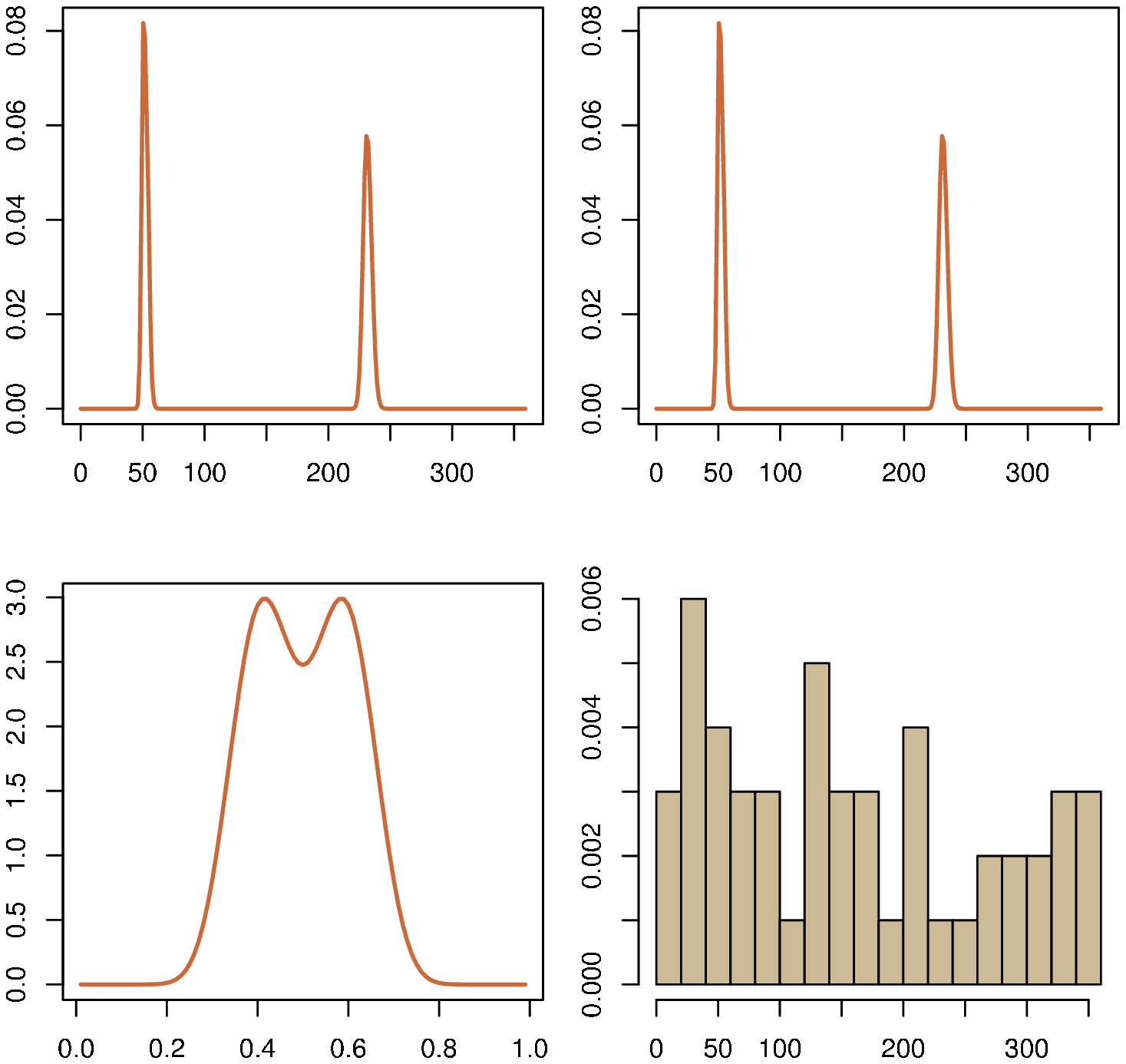}
\caption{
Same legend as Figure \ref{fig:topminnow1} for a symmetric prior
with hyperparameter $1/100$.}\label{fig:topminnow2}
\end{figure}

If we instead use a completely symmetric prior with identical hyperparameters for $\lambda_1$ and
$\lambda_2$, the output of the algorithm is then also symmetric in both components, as shown by
Figure \ref{fig:topminnow2}. The modes of the marginals of $\lambda_1$ and $\lambda_2$ remain the
same, nonetheless.

\subsection{Multinomial mixtures}\label{ex:fullM}
The case of a multinomial mixture can be dealt with similarly:
If we have $n$ observations $\bn_j=(n_{j1},\ldots,n_{jk})$ from the mixture
$$
\bn_j\sim
p\mathcal{M}_k(d_j;q_{11},\ldots,q_{1k})+(1-p)\mathcal{M}_k(d_j;q_{21},\ldots,q_{2k})\,
$$
where $n_{j1}+\cdots+n_{jk}=d_j$ and $q_{11}+\cdots+q_{1k}=q_{21}+\cdots+q_{2k}=1$, 
the conjugate priors on the $q_{ij}$s are Dirichlet distributions $(i=1,2)$,
$$
(q_{i1},\ldots,q_{ik}) \sim \mathcal{D}(\alpha_{i1},\ldots,\alpha_{ik})\,,
$$
and we use once again the uniform prior on $p$. (A default choice for the $\alpha_{ij}$'s
is $\alpha_{ij}=1/2$.) Note that the $d_j$s may differ from observation to observation, since
they are irrelevant for the posterior distribution: given a partition $\bz$ of the sample,
the complete posterior is indeed 
$$
p^{n_1} (1-p)^{n_2}\,\prod_{i=1}^2\prod_{z_j=i}q_{i1}^{n_{j1}}\cdots q_{ik}^{n_{jk}}\,
\times \prod_{i=1}^2\prod_{h=1}^k q_{ih}^{-1/2},
$$
up to a normalising constant that does not depend on $\bz$.

More generally, if we consider a mixture with $m$ components,
$$
\bn_j\sim \sum_{\ell=1}^m 
p_\ell\mathcal{M}_k(d_j;q_{\ell 1},\ldots,q_{\ell k})\,,
$$
the complete posterior is also directly available, as
$$
\prod_{i=1}^m p_i^{n_i}\times
\prod_{i=1}^m\prod_{z_j=i}q_{i1}^{n_{j1}}\cdots q_{ik}^{n_{jk}}\,
\times \prod_{i=1}^m\prod_{h=1}^k q_{ih}^{-1/2},
$$
once more up to a normalising constant.

The corresponding normalising constant of the Dirichlet distribution being
$$
K(\alpha_{i1},\ldots,\alpha_{ik}) = \frac{\prod_{j=1}^k \Gamma(\alpha_{ij}) }{\Gamma(\alpha_{i1}
+\cdots+\alpha_{ik})}\,,
$$
it produces the overall weight of a given partition $\bz$ as
\begin{equation}\label{eq:normaM}
n_1!n_2!\frac{\prod_{j=1}^k \Gamma(\alpha_{1j}+S_{1j})}{\Gamma(\alpha_{11}
+\cdots+\alpha_{1k}+S_{1\cdot})}\times\frac{\prod_{j=1}^k 
\Gamma(\alpha_{ij}+S_{2j})}{\Gamma(\alpha_{21} +\cdots+\alpha_{2k}+S_{2\cdot})}\,,
\end{equation}
where $n_i$ is the number of observations allocated to component $i$,
$S_{ij}$ is the sum of the $n_{\ell j}$s for the observations $\ell$
allocated to component $i$ and 
$$
S_{i\cdot}=\sum_j \sum_{z_\ell=i} n_{\ell j}\,.
$$ 

Given that the posterior distribution only depends on those ``sufficient" statistics 
$S_{i j}$ and $n_i$, the same factorisation as in the Poisson case applies, namely that we simply 
need to count the number of occurrences of a particular local sufficient statistic 
$(n_{1},S_{11}, \ldots,S_{km})$. The book-keeping
algorithm of \citet{fearnhead:2005} applies in this setting as well. 
What follows is a na\"ive R program translating the above:
\begin{verbatim}
em=dim(dat)[2]
emp=em+1
empcomp=emp*ncomp

#Matrix of sufficient statistics: 
#last column is number of occurrences
#each series of (em+1) columns contains, first, number of allocations 
# and, last, sum of multinomial observations
cardin=matrix(0,ncol=empcomp+1,nrow=ncomp)
\end{verbatim}

Therefore, the $(k+1)$th column of \verb+cardin+ contains the sum of the $d_j$s for
the $j$'s allocated to the first component.

\begin{verbatim}
#Initialisation
for (i in 1:ncomp) cardin[i,emp*(i-1)+(1:emp)]=c(1,dat[1,])
cardin[,empcomp+1]=1

#Update
for (i in 2:dim(dat)[1]){

    ncard=dim(cardin)[1]
    update=matrix(t(cardin),ncol=empcomp+1,nrow=ncomp*ncard,byrow=T)

    for (j in 0:(ncomp-1)){

      indi=j*ncard+(1:ncard)
      empj=emp*j
      update[indi,empj+1]=update[indi,empj+1]+1
      update[indi,empj+(2:emp)]=t(t(update[indi,empj+(2:emp)])+dat[i,])
      }

    update=update[do.call(order,data.frame(update)),]

    nu=dim(update)[1]
    #changepoints
    jj=c(1,(2:nu)[apply(abs(update[2:nu,1:empcomp]-update[1:(nu-1),
                  1:empcomp]),1,sum)>0])
    # duplicates or rather ncomplicates!
    duplicates=(1:nu)[-jj]
    if (length(duplicates)>0){

      for (dife in 1:(ncomp-1)){

        ji=jj[jj+dife<=nu]
        ii=ji[apply(abs(update[ji+dife,1:empcomp]-
                  update[ji,1:empcomp]),1,sum)==0]
        if (length(ii)>0)
         update[ii,empcomp+1]=update[ii,empcomp+1]+
                  update[ii+dife,empcomp+1]
      }
      update=update[-duplicates,]
      }

    cardin=update
    #print(sum(cardin[,2*ncomp+1])-ncomp^i)
  }
\end{verbatim}
where \verb+dat+ is now a matrix with $k$ columns.

The computation of the number of replicates of a given sufficient statistic 
$$
\mathbf{\sigma}=
(n_1,S_{11},\ldots, n_m,S_{1m},\ldots,S_{km})\,,
$$
$\mu_n(\sigma)$, is then provided by the 
last column of the matrix \verb+cardin+. The overall weight is then computed as
the product of $\mu_n(\sigma)$ with the normalising constant \eqref{eq:normaM}:
\begin{verbatim}
olsums=matrix(0,ncol=ncomp,nrow=dim(update)[1])

for (y in 1:ncomp)
 colsums[,y]=apply(update[,(y-1)*emp+(2:emp)],1,sum)

w=log(cardin[,empcomp+1])+
    apply(lfactorial(cardin[,emp*(0:(ncomp-1))+1]),1,sum)+
    apply(lfactorial(cardin[,
            (1:empcomp)[-1-emp*(0:(ncomp-1))]]-.5),1,sum)-
    apply(lfactorial(colsums)+em*.5-1,1,sum)- sum(lfactorial(dat))
w=exp(w-max(w))
cardin=cbind(cardin,w)
\end{verbatim}
As shown in Table \ref{tab:multi},
once again, the reduction in the number of cases to be considered is enormous.

\begin{table}\begin{center}
\begin{tabular}{l cccc}
\toprule
$(n,d_j,k)$   & $m=2$ & $m=3$ & $m=4$ & $m=5$\\
\midrule
$(10,5,3)$  &$33/35$ &$4602/9093$ &$56,815/68,964$ &--\\ 
$(10,5,4)$  &$90/232$ &$3650/21,249$ &$296,994/608,992$ &--\\ 
$(10,5,5)$  &$247/707$	&$247/7857$ &$59,195/409,600$ &-- \\
$(10,10,2)$ &$19/20$ &$803/885$ &$3703/4800$ &$7267/11550$\\
$(10,10,3)$ &$117/132$ &$1682/1893$ &$48,571/60,720$ &--\\
$(10,10,4)$ &$391/514$ &$3022/3510$ &$83,757/170,864$ &--\\
$(10,10,5)$ &$287/1008$ &$7,031/12,960$ &$35,531/312,320$ &--\\
$(20,5,2)$ &$129/139$ &$517/1140$ &$26,997/45,600$ &$947/10,626$\\
$(20,5,3)$ &$384/424$ &$188,703/209,736$ &$108,545/220,320$ &--\\
$(20,5,4)$ &$3410/6944$ &$819,523/1,058,193$ &-- &--\\
$(20,10,5)$ &$1225/1332$ &$9,510/1,089,990$ &-- &--\\
\botrule
\end{tabular}
\caption{\label{tab:multi}
Number of sufficient statistics $(n_i,S_{ij})_{1\le i\le m,1\le j\le k}$ corresponding to the $99$\%~largest posterior,
of pairs $(n_i,S_i)$ corresponding to the $99$\%~largest posterior weights, and total number of statistics
for datasets simulated from mixtures of $m$ multinomial $\mathcal{M}_k(d_j;q_1,\ldots,q_k)$ 
and different parameters.}{\em (Missing terms are due
to excessive computational or storage requirements.)}
\end{center}\end{table}

\subsection{Normal mixtures}\label{sub:fullN}

For a normal mixture, the number of truly different terms in the posterior distribution is much
larger than in the previous (discrete) cases, in the sense that only permutations of the members of
a given partition within each term of the partition provide the same local sufficient statistics.
Therefore, the number of observations that can be handled in an exact analysis is necessarily
extremely limited. 

As mentioned in Section \ref{sub:Locconj}, the locally conjugate priors for normal
mixtures are products of normal $\mu_j|\sigma_j \sim \mathcal{N}(\xi_j,\sigma^2_j/c_j)$
by inverse gamma $\sigma^{-2}_j\sim \mathcal{G}(a_j/2,b_j/2)$ distributions. For instance,
in the case of a two-component normal mixture,
$$
x_1,\ldots,x_n\stackrel{\text{iid}}{\sim}
p\,\mathcal{N}(\mu_1,\sigma_1^2) + (1-p)\,\mathcal{N}(\mu_2,\sigma_2^2)\,,
$$
we can pick 
$\mu_1|\sigma_1 \sim \mathcal{N}(0,10\sigma^2_j)$, 
$\mu_2|\sigma_2 \sim \mathcal{N}(1,10\sigma^2_2)$,
$\sigma^{-2}_j\sim \mathcal{G}(2,2/\sigma_0^2)$,
if a difference of one between both means is considered likely (meaning
of course that the data are previously scaled) and if $\sigma_0^2$ is the prior 
assumption on the variance (possibly deduced from the range of the sample).
Obviously, the choice of a Gamma distribution with $2$ degrees of freedom is
open to discussion, as it is not without consequences on the posterior distribution.

The normalising constant of the prior distribution is (up to a true constant)
$$
K(a_1,\ldots,a_k,b_1,\ldots,b_k,c_1,\ldots,c_k,\xi_1,\ldots,\xi_k) =
\prod_{i=1}^k \sqrt{c_j}\,.
$$
Indeed, the corresponding posterior is
$$
\mu_j|\sigma_j \sim \mathcal{N}\left[(c_j\xi_j+n_j\overline{x}_j,\sigma^2_j/(c_j+n_j)\right]
$$
and
$$
\sigma^{-2}_j\sim \mathcal{G}\left[\{a_j+n_j\}/2,\{b_j+n_j\hat\sigma^2_j
        +(\overline{x}_j-\xi_j)^2/(c_j^{-1}+n_j^{-1})\}\right]\,.
$$
The number of different sufficient statistics $(n_j,\overline{x}_j,\hat\sigma^2_j)$ is thus related
to the number of different partitions of the dataset into at most $k$ groups. This is related to
the Bell number \citep{rota:1964}, which grows extremely fast. We therefore do not pursue the example
of the normal mixture any further for lack of practical purpose. 

\section*{Acknowledgements} 
This paper is a chapter of the book {\em Mixtures:
Estimation and Applications}, edited by the authors jointly with Mike
Titterington and following the ICMS workshop on the same topic that took place
in Edinburgh, March 03-05, 2010. The authors are deeply grateful to the staff
at ICMS for the organisation of the workshop, to the funding bodies (EPSRC,
LMS, Edinburgh Mathematical Society, Glasgow Mathematical Journal Trust, and
Royal Statistical Society) for supporting this workshop, and to the
participants in the workshop for their innovative and exciting contributions.


\end{document}